\documentclass[english,reprint,amsmath,amssymb,aps,prl,showpacs]{revtex4-1}
\usepackage[T1]{fontenc}
\usepackage[latin9]{inputenc}
\usepackage{graphicx}
\usepackage{babel}
\begin{document}

\title{Time Evolution of Quantum Entanglement of an EPR Pair in a Localized
Environment}

\author{Jia Wang}

\affiliation{Centre for Quantum and Optical Science, Swinburne University of Technology,
Melbourne 3122, Australia}

\author{Xia-Ji Liu}

\affiliation{Centre for Quantum and Optical Science, Swinburne University of Technology,
Melbourne 3122, Australia}

\author{Hui Hu}

\affiliation{Centre for Quantum and Optical Science, Swinburne University of Technology,
Melbourne 3122, Australia}

\date{\today}
\begin{abstract}
The Einstein\textendash Podolsky\textendash Rosen (EPR) pair of qubits
plays a critical role in many quantum protocol applications such as
quantum communication and quantum teleportation. Due to interaction
with the environment, an EPR pair might lose its entanglement and
can no longer serve as useful quantum resources. On the other hand,
it has been suggested that introducing disorder into environment might
help to prevent thermalization and improve the preservation of entanglement.
Here, we theoretically investigate the time evolution of quantum entanglement
of an EPR pair in a random-field XXZ spin chain model in the Anderson
localized (AL) and many-body localized (MBL) phase. We find that the
entanglement between the qubits decreases and approaches to a plateau
in the AL phase, but shows a power-law decrease after some critical
time determined by the interaction strength in the MBL phase. Our
findings, on one hand, shed lights on applying AL/MBL to improve quantum
information storage; on the other hand, can be used as a practical
indicator to distinguish the AL and MBL phase. 
\end{abstract}
\maketitle
An Einstein-Podolsky-Rosen (EPR) pair is a pair of qubits which are
in maximally entangled state. Due to their perfect quantum correlations,
EPR pairs lie at the heart of many important proposals for quantum
communication and computation, such as quantum teleportation \cite{Ekert1991,Bennett1993}.
In reality, however, due to the unavoidable decoherence induced by
the couplings to the surrounding environment, an EPR pair might become
a state $\rho$ that lose entanglement after a certain time, making
this qubit pair no longer useful as a quantum resource. One of the
main tool to overcome the decoherence is a protocol named entanglement
distillation \cite{Bennett1996,Horodecki1998}. This method can be
used to transform $N$ copies of less entangled states $\rho$ back
into a smaller number $m$ of approximately pure EPR pairs by using
only local operations and classical communication (LOCC), where the
ratio $m/N$ depends on the amount of entanglement left in $\rho$.
Therefore, it is of great interest to design quantum information storage
devices that can keep a strong quantum entanglement for a long time
to improve the distillation efficiency. In this Letter, we study the
possibility of preserving quantum entanglement in a localized environment
by introducing strong disorder. 

The idea that disorder can help protecting initial correlations and
information is first raised by Anderson in 1958. He focused on the
behaviors of non-interacting particles experiencing random potentials,
which is now named as Anderson localization (AL) \cite{Anderson1958}.
In AL, the diffusion of particle's wave-packet in a disordered environment
is absent, implying the initial information of particle's position
is ``remembered''. Extending this concept to an interacting system,
namely many-body localization (MBL), has attracted many people's interest
including Anderson himself. Recently, this field attracts an intense
attraction \cite{Basko2006,Altman2015,Nandkishore2015}, partially
due to the lately rapid progress in ultracold atomic experiments that
has made quantum isolated many-body systems with tunable interaction
and disorder available, including ultracold atoms in optical lattices
\cite{Schreiber2015,Bordia2016,Choi2016} and ion traps \cite{Smith2015}.
These experimentally available systems constitute promising platforms
for exploring the AL and MBL localization phases and stimulated a
series of theoretical studies. Many remarkable properties of these
localized phases have ever since been theoretically predicted: Poisson
distributions of energy gap \cite{Oganesyan2007,Atas2013}, absences
of transportation of charge, spin, mass or energy even at high temperature
\cite{Berkelbach2010,Pekker2014,Agarwal2015}, protecting quantum
order and discrete symmetry that normally only exists in the ground
state \cite{Huse2013,Bahri2013,Bauer2013,Chandran2014} and existence
of mobility edge \cite{Luitz2015,Li2015,Baygan2015}. In particular,
quantum entanglement has been discovered to play very important roles
in identifying different phases: energy eigenstates in localized phases
have area-law bipartition entanglement entropy in contrast to the
volume-law entropy of a thermalized state \cite{Znidari2008,Kajall2014}.
In addition, after a sudden (global or local) quench, the entanglement
shows a fast power-law spreading in a thermal phase, but only a slow
logarithmic spreading in an MBL phase and no spreading at all in an
AL phase \cite{Bardarson2012,Serbyn2013,Deng2016}. The slow entanglement
spreading in the localized phases is restricted by a variant of the
Lieb-Robinson bound on the information light cone, which can in principle
be observed via out-of-time-order correlations (OTOC) \cite{Huang2016,Fan2016,Chen2016}.
This slow spreading of entanglement also suggests that the local correlations
might be maintained for a long time in a localized environment, which
has a potential application in quantum information storage. Indeed,
it has been shown that deep in the localized phase, the quantum coherence
of local degrees of freedom, e.g. a single qubit, has been demonstrated
to be maintained for a very long time \cite{Serbyn2014,Bahri2015,Vasseur2015}.
However, to the best of our knowledge, whether disorder can also help
to protect quantum entanglement between qubits has never been directly
studied. In this Letter, we focus on studying the time evolution of
the quantum entanglement between an EPR pair shared by two observers
namely Alice and Bob and coupled to a localized environment.

As a concrete example, we conduct our analysis in a prototype Hamiltonian
that has been studied extensively in the MBL literature: a one-dimensional
(1D) $s = 1/2$ spin chain XXZ Hamiltonian with nearest neighbor interactions

\begin{equation}
\ensuremath{H=\sum\limits _{i}{J\left({s_{i}^{x}s_{i+1}^{x}+s_{i}^{y}s_{i+1}^{y}}\right)+\Delta s_{i}^{z}s_{i+1}^{z}+{h_{i}}s_{i}^{z}}},\label{eq:Hamiltonian}
\end{equation}
where $J$ and $\Delta$ are both constant, and $h_i$ are random
fields uniformly distributed over $[-h,h]$. The total magnetization
${S_z} \equiv \sum\limits_i {s_i^z}$ is a good quantum number, and
hence we will restrict our calculation for $S_z=0$ hereafter. We
want to emphasize here that the spin-spin interacting Hamiltonian
is chosen not only because its localized phase has been well studied,
but also because in some reality cases, the main source of decoherence
for qubits are from their interaction with unwanted environment spins.
We also remark here that, the Hamiltonian in Eq. (\ref{eq:Hamiltonian})
can be mapped into a Fermi-Hubbard model using a Jordan-Wigner transformation,
where $J$ is equivalent to the hopping coefficient and $\Delta$
is equivalent to the interaction strength. Thus with strong enough
disorder $h$, the spin chain is expected to be in the AL (MBL) phase
for $\Delta=0$ ($\Delta \ne 0$) respectively. 

Here, we prepare an initial state in the form of $\left| {\Psi \left( 0 \right)} \right\rangle ={\left| {\rm EPR} \right\rangle _{\rm AB}} \otimes {\left| {\rm NEEL} \right\rangle _E}$,
where the subscript A stands for Alice's spin, B stands for Bob's,
and E stands for all the other spins serving as an environment. The
environment is prepared in a N{\'e}el's state mimicking a high-temperature
environment and is initially \emph{not} entangled with the EPR pair
of the spin A and B. We then study the time evolution of this state
under the Hamiltonian in Eq. (\ref{eq:Hamiltonian}) using exact diagonalization,
obtaining the reduced density matrix $\rho$ of the spin pair A and
B by tracing out all the environment spins and calculating quantum
entanglement measurements that are usually averaged over many realizations
of disorder (typically 1000 times). The quantum entanglement measurement
we focus here is the logarithmic negativity is given by $S_N=\log_2\left(N+1\right)$,
where $N$, namely negativity, is a measure related to the Peres-Horodecki
criterion: $N = 2\sum\nolimits_i {\max \left( {0, - {\mu _i}} \right)}$
\cite{Peres1996,Horodecki1996}. Here $\mu_i$'s are the eigenvalues
of $\rho^{\Lambda}$ who is the partial transpose of $\rho$. We would
like to remark here that, the logarithmic negativity, even though
lacks convexity, is a full entanglement monotone that does not increase
on average under a general positive partial transpose (PPT) preserving
operation as well as local operations and classical communication
(LOCC) \cite{Plenio2005}. In addition, the logarithmic negativity
serves as the upper bound of distillable entanglement that limits
the amount of nearly maximally entangled qubit pairs that can be asymptotically
distilled from $N$ copies of $\rho$ via quantum distillation \cite{Bennett1996,Horodecki1998}. 

\begin{figure}
\includegraphics[width=9cm]{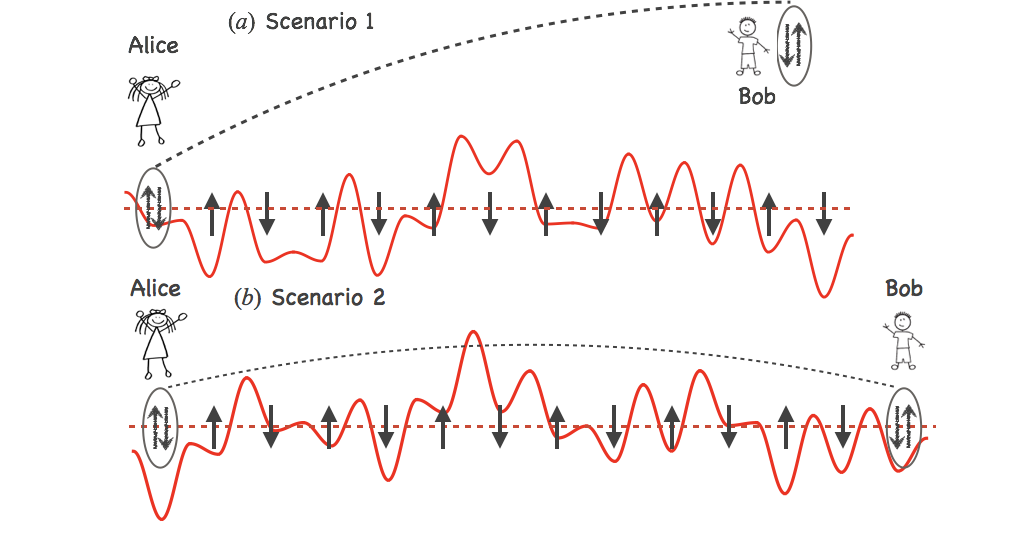}\caption{A sketch of the two scenarios that studied here. (a) The first scenario,
where both Alice and Bob's spins are immersed in the same disordered
environment. (b) The second scenario, where Bob is isolated from the
environment that Alice experiences.\label{fig:Scenarios}}
\end{figure}

In our current set-up, two scenarios can be studied: in the first
scenario shown in Fig. \ref{fig:Scenarios}(a), Bob is isolated from
the environment, which resembles a quantum communication or quantum
teleportation situation; in the second scenario illustrated in Fig.
\ref{fig:Scenarios}(b), both Alice and Bob are in contact with the
same environment mimicking a quantum calculation realization. Our
numerical result shows that the entanglement evolutions in both scenarios
have similar qualitative behavior. Therefore, we focus on discussing
the logarithmic negativity for scenario one here. These discussion
and conclusions are however applicable for other entanglement measurements
such as concurrence and entanglement of formation in both scenarios
\cite{Supplemental}.

\begin{figure}
\includegraphics[width=9cm]{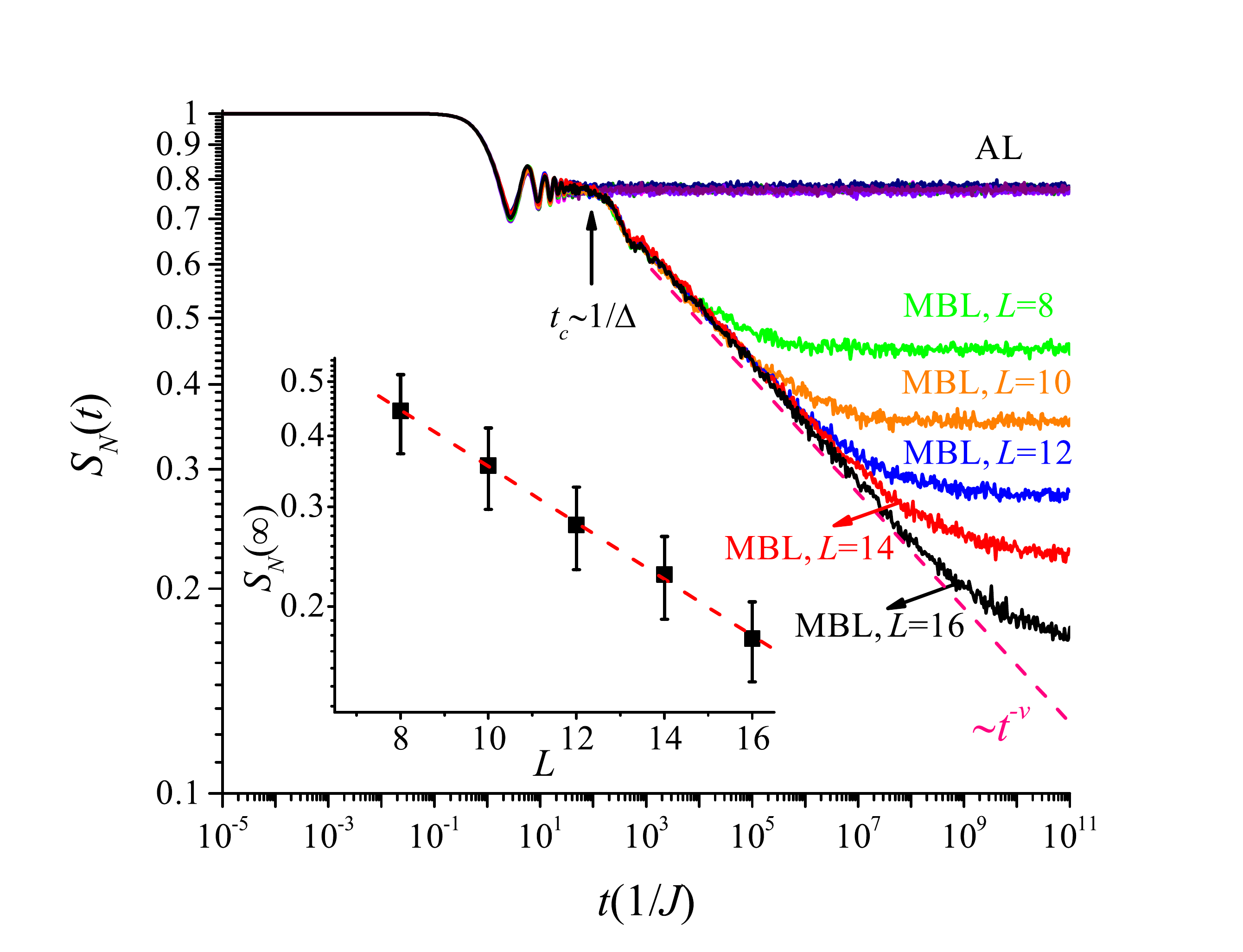}\caption{Logarithmic negativity in AL $(\Delta=0J)$ and MBL $(\Delta=10^{-2}J)$
phases as a function of time for disorder strength $h=3$ and different
spin chain length $L$. The dashed line helps to highlight the power-law
decay behavior in the MBL phases. The inset shows the final saturation
values of entanglement $S_N(\infty)$ in MBL phases as a function
of $L$, where the error bar shows the variance from averaging over
different disorder realizations. The dashed line shows a exponential
fit of $S_N(\infty)\sim \exp(-L)$. \label{fig:SNvsTime}}
\end{figure}

\begin{figure}
\includegraphics[width=9cm]{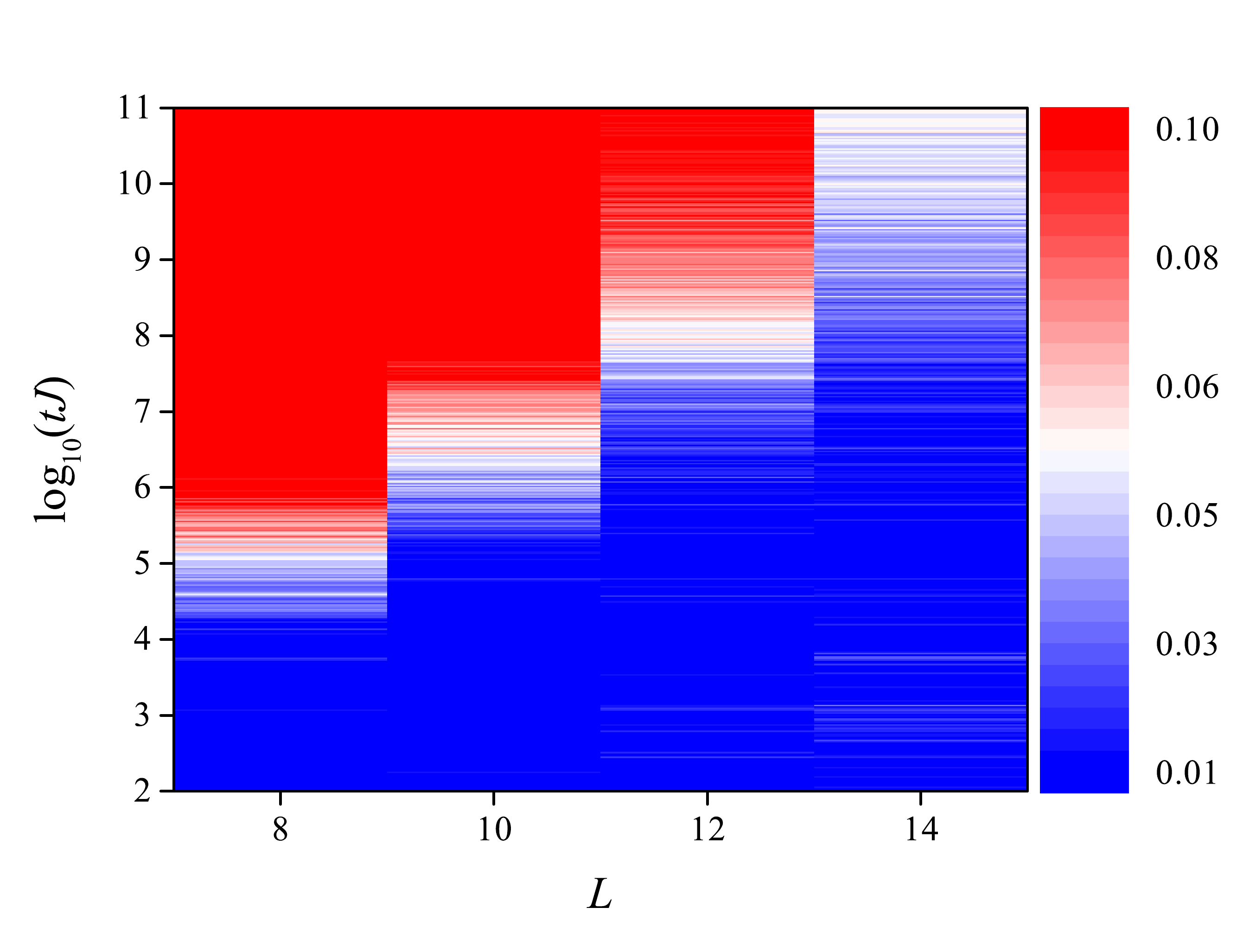}\caption{Color plot of $\Delta S_N^{(L)}$ defined in the main text. The red
color indicates a large value of $\Delta S_N^{(L)}$, which is constrained
by a logarithmic light cone.\label{fig:DeltS}}
\end{figure}

Figure \ref{fig:SNvsTime} shows our main result, the logarithmic
negativity between Alice and Bob as a function of time in the scenario
one, with the disorder strength $h=3 J$ for different numbers of
spins $L$ (including Alice and Bob's spins) and $\Delta=0\ (10^{-2})$
for the AL (MBL) phases. Initially, the entanglement is prepared at
maximum $S_N(0)=1$. At around $t \approx 1/J$, the entanglement
in both AL and MBL phases shows a power-law decay following some oscillations,
which has been recognized as the diffusion of initial state to a size
of the localization length. After about a critical time $t_c \approx 1/\Delta$,
the entanglement of MBL and AL shows dramatically different behavior.
The entanglement in AL phases converges to a plateau independent of
the spin chain size $L$, where all curves for different $L$ are
visually overlapping. In contrast, the entanglement in MBL phases
shows a power law decay $\sim {t^{ - v}}$ with $v>0$, which is emphasized
by the linear behavior on a log-log scale in Fig. \ref{fig:SNvsTime}.

Due to the finite size of our system, the entanglement in MBL phases
will eventually also saturate to some constants after a very long
time. Nevertheless, as illustrated in the inset of Fig. \ref{fig:SNvsTime},
the final saturated values are shown to decrease exponentially as
a function of spin chain size $\sim \exp \left( { - \beta L } \right)$,
where $\beta$ is a constant. From these observations, one can expect
that the entanglement in AL phase will never reduce to zero, but a
constant depend only on disorder strength even in the thermodynamic
limit $L \rightarrow \infty$. On the other hand, no matter how small
the interaction strength $\Delta$ is, the entanglement will be completely
dissipated after infinite long time in the thermodynamic limit. However,
this dissipation is very slow if the disorder is strong enough. In
addition, the entanglement of AL phase and MBL phase only become different
abruptly after the critical time $t_c$, therefore, if $\Delta$ is
small, the entanglement can still be preserved in the AL level before
$t_c$. 

Therefore, we can conclude that the AL phase is ideal for creating
quantum storage devices to preserve quantum entanglement between a
qubit pair. On the other hand, if a weak interaction strength $\Delta$
is unavoidable in the system, the MBL phase can still be applied to
preserve entanglement but with an expiration time $t_c$. However,
one must carry an entanglement distillation to use the qubit pair
before conducting quantum protocols. We also wish to emphasize here
that the preservation of entanglement between two qubits in a localized
environment is, of course, not better than in a completely decoupled
environment. However, in a realistic situation where coupling between
the qubit pair and the environment cannot be eliminated, our study
provides a generic way of preserving entanglement without a specific
fine tuning of the Hamiltonian but simply introducing strong enough
disorder into the environment. 

Our results can also be directly applied to identify the localized
phase being AL or MBL. Most of previous such studies have been focused
on studying the bipartite entanglement (, i.e. dividing the system
into two sub-systems,) of an initial product state after global quench
or an energy eigenstate after a local quench \cite{Bardarson2012,Serbyn2013,Deng2016}.
Nevertheless, the experimental observation of bipartite entanglement
in principle can be very challenging for a large system and may even
be impossible in the thermodynamic limit. On the other hand, measuring
entanglement between local degrees of freedom in an optical lattice
\cite{Fukuhara2015} and trap ions \cite{Jurcevic2014,Esteban2016}
has been reported lately. Therefore, studies of entanglement between
two sites have been investigated recently, motivated by the fact that
such entanglement between local degrees of freedom is much more experimentally
accessible \cite{Bera2016,Tomasi2016,Iemini2016}. These studies usually
focus on the case where the initial state is a product state, and
a temporary entanglement generated due to initial diffusion. The AL
and MBL features are analyzed by the following decay of this temporary
entanglement that decreases exponentially as a function of distances
between sites in the deep localized phase. Therefore, these studies
are usually limited to entanglement between nearest few sites. Our
methods, using an initial prepared EPR pair, can in principle overcomes
these limitations and be experimentally accessible. 

Our study also gives an interesting insight into the nature of entanglement
spread in MBL phases. The power law decay and the saturated values
of entanglement in MBL phases suggest that we can define a saturation
time scale $t_s$, where the entanglement in MBL phases is about to
be saturated, as $v\log \left( t_s \right) \sim L$, which resembles
the logarithmic light-cone found in previous bipartite entanglement
studies \cite{Bardarson2012,Serbyn2013,Deng2016}. This logarithmic
light-cone can be understood from the modified Lieb-Robinson bound
of information spreading (in this case entanglement spreading) \cite{Friesdorf2015}.
One convenient and common way to describe the Lieb-Robinson bound
is to compare the time-evolution of a local observable $A$ under
the full Hamiltonian with its time-evolution under a truncated Hamiltonian
that only includes interactions contained in a region of distance
no more than $L$. Even though entanglement is technically not an
observable, we study the quantity $\Delta S_N^{(L)}=S_N^{(L=16)}-S_N^{(L)}$
under the same spirit. In scenario two, this quantity can be interpreted
as the differences of entanglement between a full spin chain of $L=16$
and a truncated spin chain $L$ near Alice's spin. The result is shown
in Fig. \ref{fig:DeltS}, where one can directly see that the significant
differences are constrained within a logarithmic light cone. This
is a direct evidence that entanglement is spreading logarithmically
in an MBL phase suggested by previous OTOC studies \cite{Huang2016,Fan2016,Chen2016}.

\begin{figure}
\includegraphics[width=9cm]{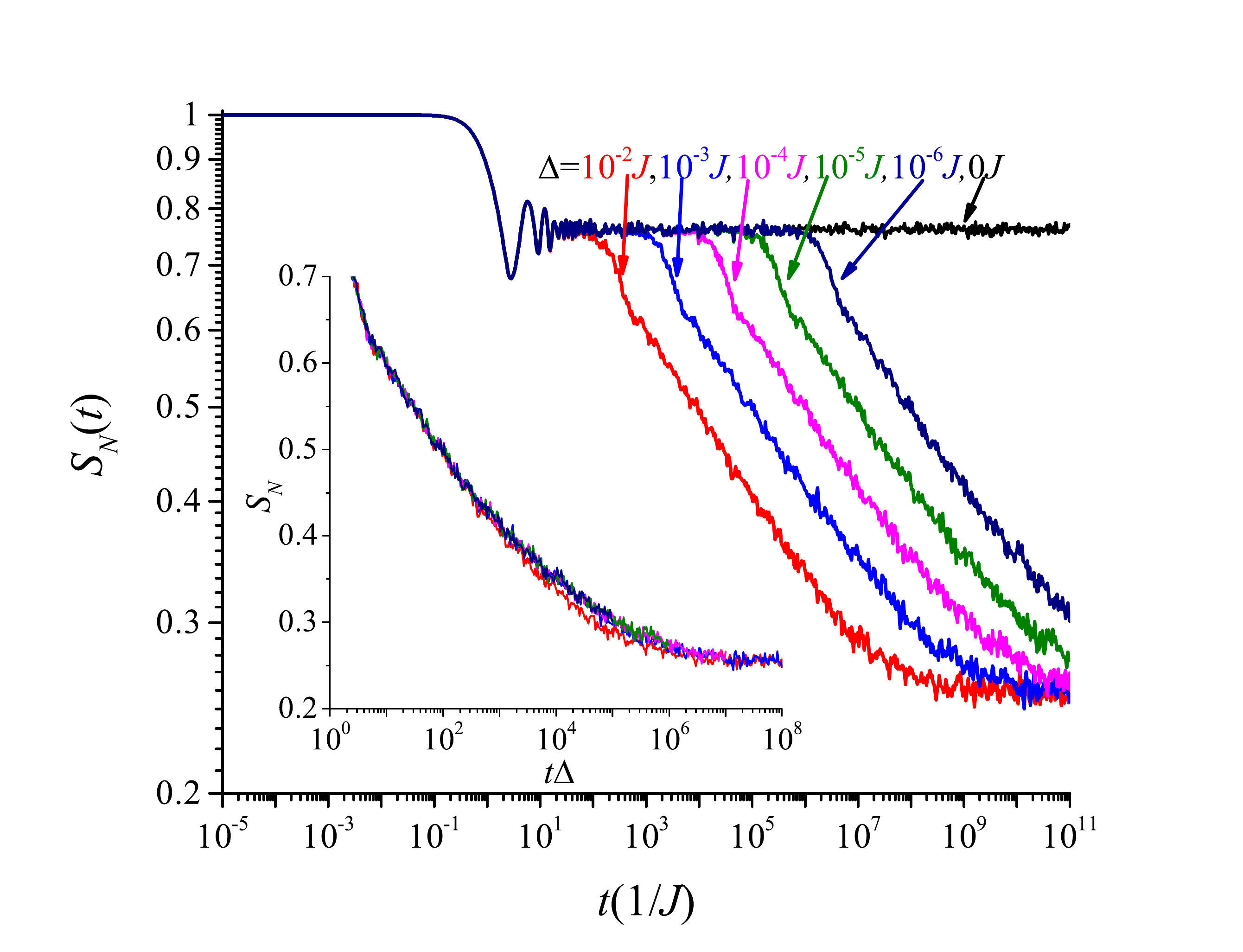}\caption{Entanglement decay for different interaction strength $\Delta$. Here,
the size of spin chain $L=12$ and $h=3 J$. The inset shows the long-time
behavior of entanglement as a universal function of $t\Delta$. \label{fig:D-dependent}}
\end{figure}

\begin{figure}
\includegraphics[width=9cm]{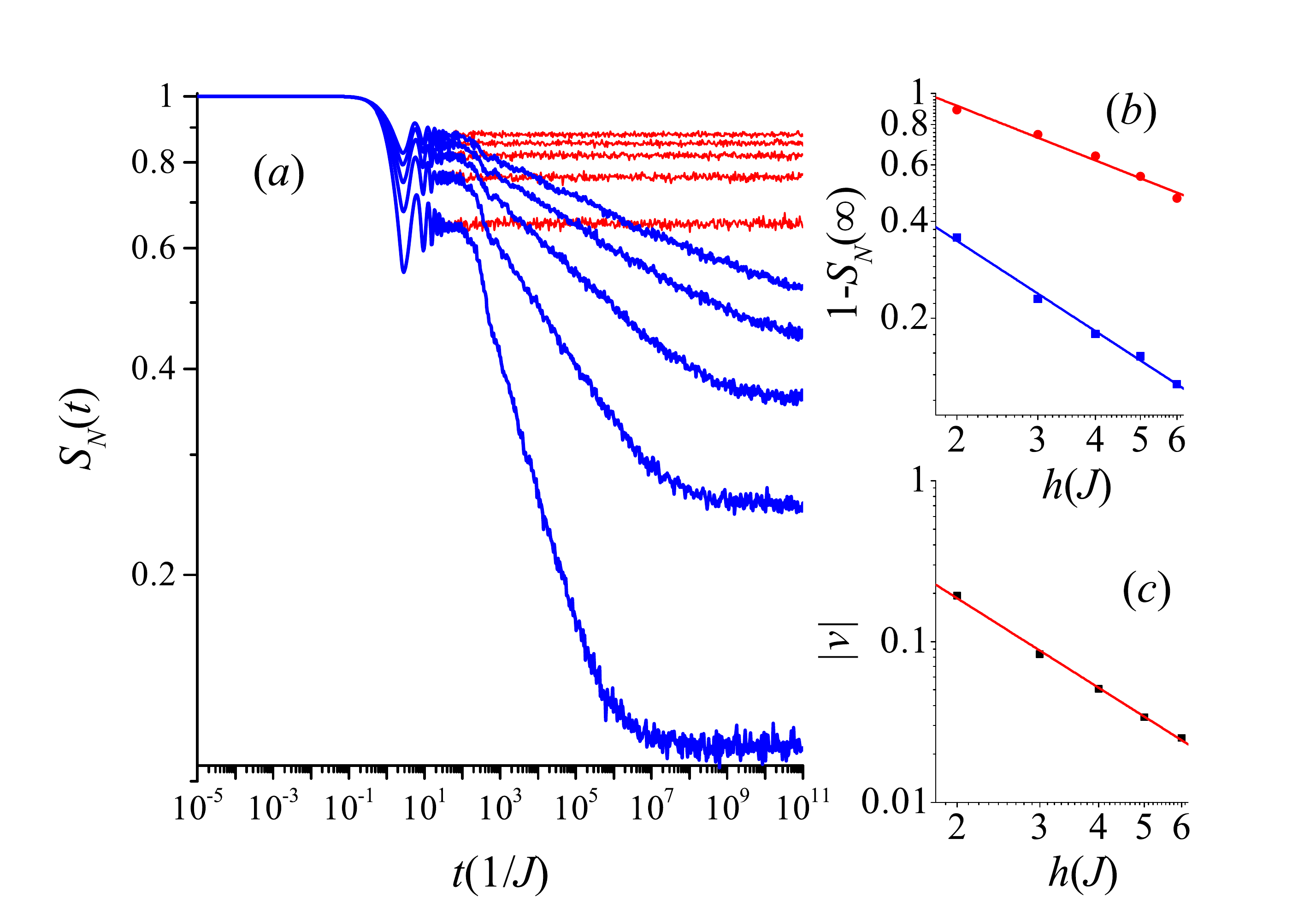}\caption{(a) Entanglement decay for different disorder strength $h$ with $L=12$
and $\Delta=10^{-2} J$. The red thin (blue thick) curves represents
AL (MBL) phases for $h=2J,3J,4J,5J,6J$ from bottom to top. (b) The
saturated value $1-S_N(\infty)$ as a function of $h$, the red circles
(blue squares) represent AL (MBL) phase, wher the solid lines are
power-law fit $1-S_N(\infty) \sim h^c$ . (c) The symbols are the
decay index $v$ as a function of $h$, and the solid line is a power-law
fit. \label{fig:h-dependent}}
\end{figure}

We further take a qualitative analysis of the effects of interaction
strength $\Delta$ and disorder strength $h$ on the decay of entanglement.
Figure \ref{fig:D-dependent} shows the entanglement decay for different
$\Delta$, confirming that entanglement in AL and MBL phases only
become different abruptly after the critical time $t_c \approx 1/\Delta$.
Furthermore, the saturated value in the MBL phases does not variate
appreciably for different $\Delta$. In fact, the logarithmic entanglement
for $t \gg t_c$ is a universal function of $t\Delta$ as evidenced
by the inset of Fig. \ref{fig:D-dependent}. Finally, Fig. \ref{fig:h-dependent}(a)
shows the entanglement decay for different $h$, where the decay rate
becomes slower and the saturated value becomes larger for a stronger
disorder. As a result of the competition of these two effects, the
saturation time becomes longer, suggesting that a stronger disorder
is beneficial for storing EPR pairs. Our numerical results also show
that the decrease of entanglement at infinite long time $1-S_N(\infty)$
and the decay index $v$ are both has a power-law dependence on $h$,
as shown in Fig. \ref{fig:h-dependent} (b) and (c). This analysis
can be interpreted as a stronger disorder and weaker interaction is
beneficial for preserving quantum entanglement, which is consistent
with our expectation.

In summary, we studied the time evolution of quantum entanglement
of an EPR pair coupling to a localization environment. This study
allows us to explored the possibility and limitation of applying localization
phase to preserve quantum entanglement between qubit pairs. Our results
can also be regarded as an experimentally accessible protocol to discriminate
AL and MBL phases, and understand the nature of entanglement propagation
in these systems.

This research was supported under Australian Research Council\textquoteright s
Future Fellowships funding scheme (project number FT140100003) and
Discovery Projects funding scheme (project number DP170104008). The
numerical calculations were partly performed using Swinburne new high-performance
computing resources (Green II).

\bibliographystyle{apsrev4-1}
\bibliography{EPRinMBL}

\newpage
\section*{Supplemental Material}
It is well known that von Neumann entropy is not a valid entanglement
measurement if the collective states are in mix states. However, several
entanglement measurement have been found for a pair of qubits, including
concurrence, negativity and their close relatives: entropy of formation
and logarithmic negativity. These entanglement measurements are ``good''
in the following sense: (i) for a maximally entangled state, i.e.
an EPR pair, these measurements reach their maximum values (equal
one in our definitions); (ii) for collective seperable states, these
measurements vanish; (iii) is a continuous function of density matrices
of the two-qubit states. Notice that these measurements, however,
does not nessesarily give same ordering for different entangled states. 

Let us first give the defination of the entanglement measurements
mentioned above. Denoting the collective state for two selected qubits
by a density matrix $\rho$, the concurrence is given by $C=\max(\sqrt{\lambda_1}-\sqrt{\lambda_2}-\sqrt{\lambda_3}-\sqrt{\lambda_4})$,
where $\lambda_i$'s are the eigenvalues of $\rho\tilde\rho$ in decending
order and $\tilde \rho  \equiv {\sigma ^y} \otimes {\sigma ^y}{\rho ^*}{\sigma ^y} \otimes {\sigma ^y}$.
Concurence is monotonically related to the entangelment of formation
by the Wootters formula ${S_F} = h\left[ {\left( {1 + \sqrt {1 - {C^2}} } \right)/2} \right]$,
where $h\left( x \right) =  - x{\log _2}x - \left( {1 - x} \right){\log _2}\left( {1 - x} \right)$.
Negativity is a measure related to the Peres-Horodecki criterion:
$N = 2\sum\nolimits_i {\max \left( {0, - {\mu _i}} \right)}$, where
$\mu_i$'s are the eigenvalues of $\rho^{\Lambda}$ who is the partial
transpose of $\rho$. The logarithmic negativity is then given by
$S_N=\log_2\left(N+1\right)$.

In the main text, we show that the time evolution of logarithmic negativity
between a pair of qubits that initially prepared to be as an EPR pair
in scenario one. Here, we present results of other entanglement measures
in scenario two and show that these measurements are qualitatively
similar, and serve the same role in our analysis. Therefore, the discussions
and conclusions of logarithmic negativity in scenario one are also
applicable for all measurements in scenario two. 

\begin{figure}
\includegraphics[width=9cm]{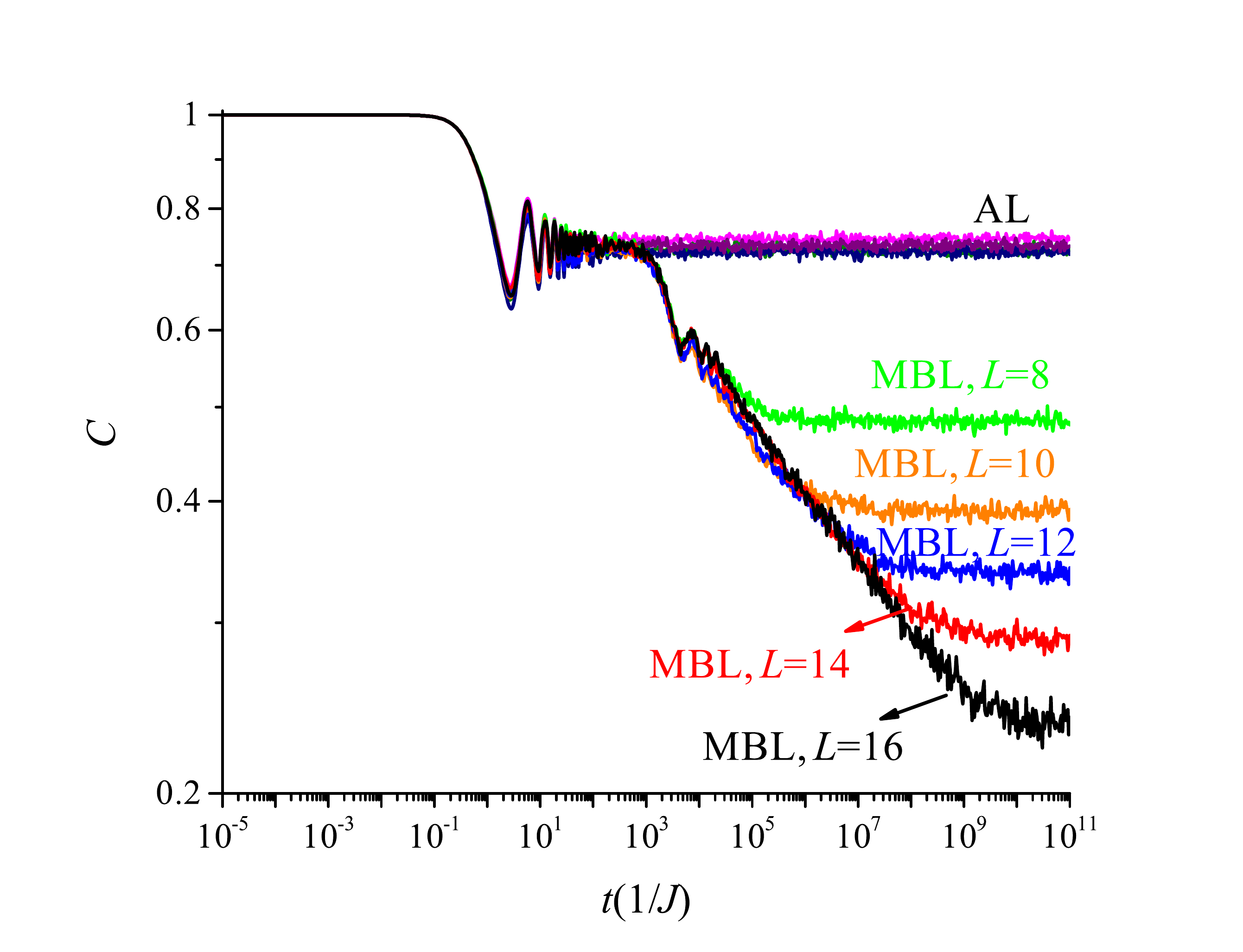}\caption{Concurrence as a function of time for disorder strength $h=5$ and
interaction strength $\Delta=0J (10^{-3}J)$ for AL (MBL) phases.
The results for AL phases with different spin chain length $L$ are
almost overlapping each other, and the results for different MBL phases
are indicated in the figure.}
\end{figure}

\begin{figure}
\includegraphics[width=9cm]{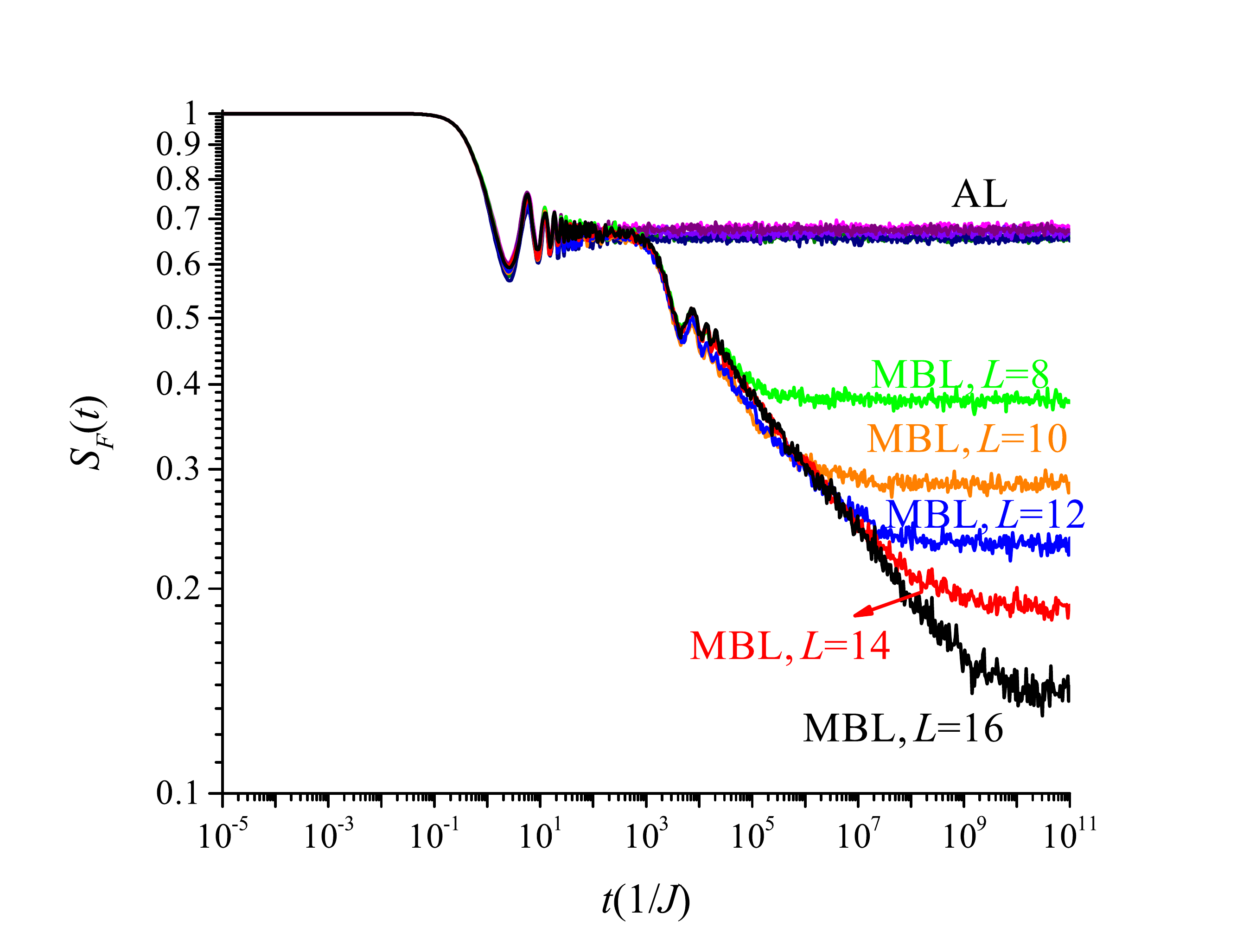}\caption{Entropy of formation as a function of time for disorder strength $h=5$
and interaction strength $\Delta=0J (10^{-3}J)$ for AL (MBL) phases.
The results for AL phases with different spin chain length $L$ are
almost overlapping each other, and the results for different MBL phases
are indicated in the figure.}
\end{figure}

\begin{figure}
\includegraphics[width=9cm]{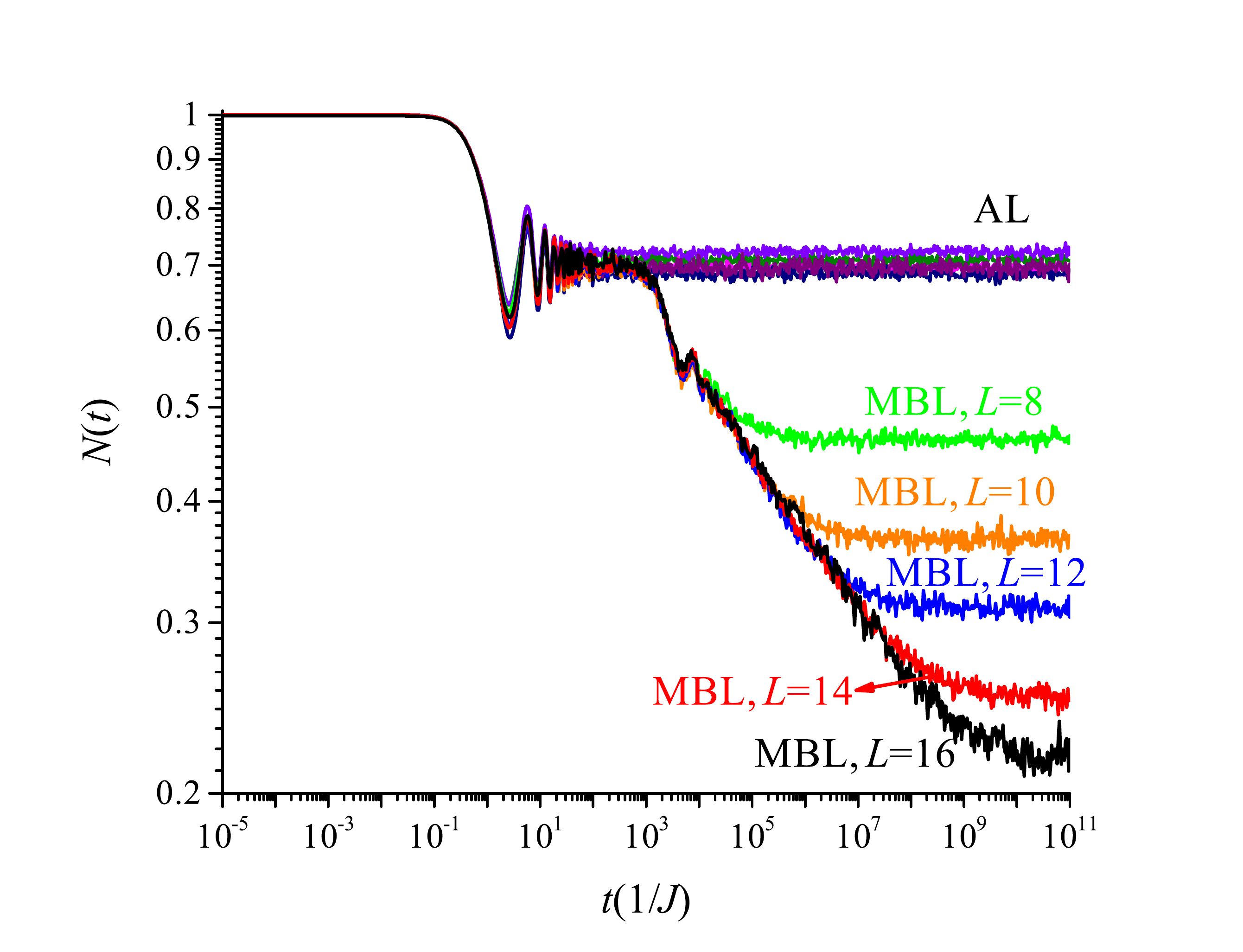}\caption{Negativity s a function of time for disorder strength $h=5$ and interaction
strength $\Delta=0J (10^{-3}J)$ for AL (MBL) phases. The results
for AL phases with different spin chain length $L$ are almost overlapping
each other, and the results for different MBL phases are indicated
in the figure.}
\end{figure}

\begin{figure}
\includegraphics[width=9cm]{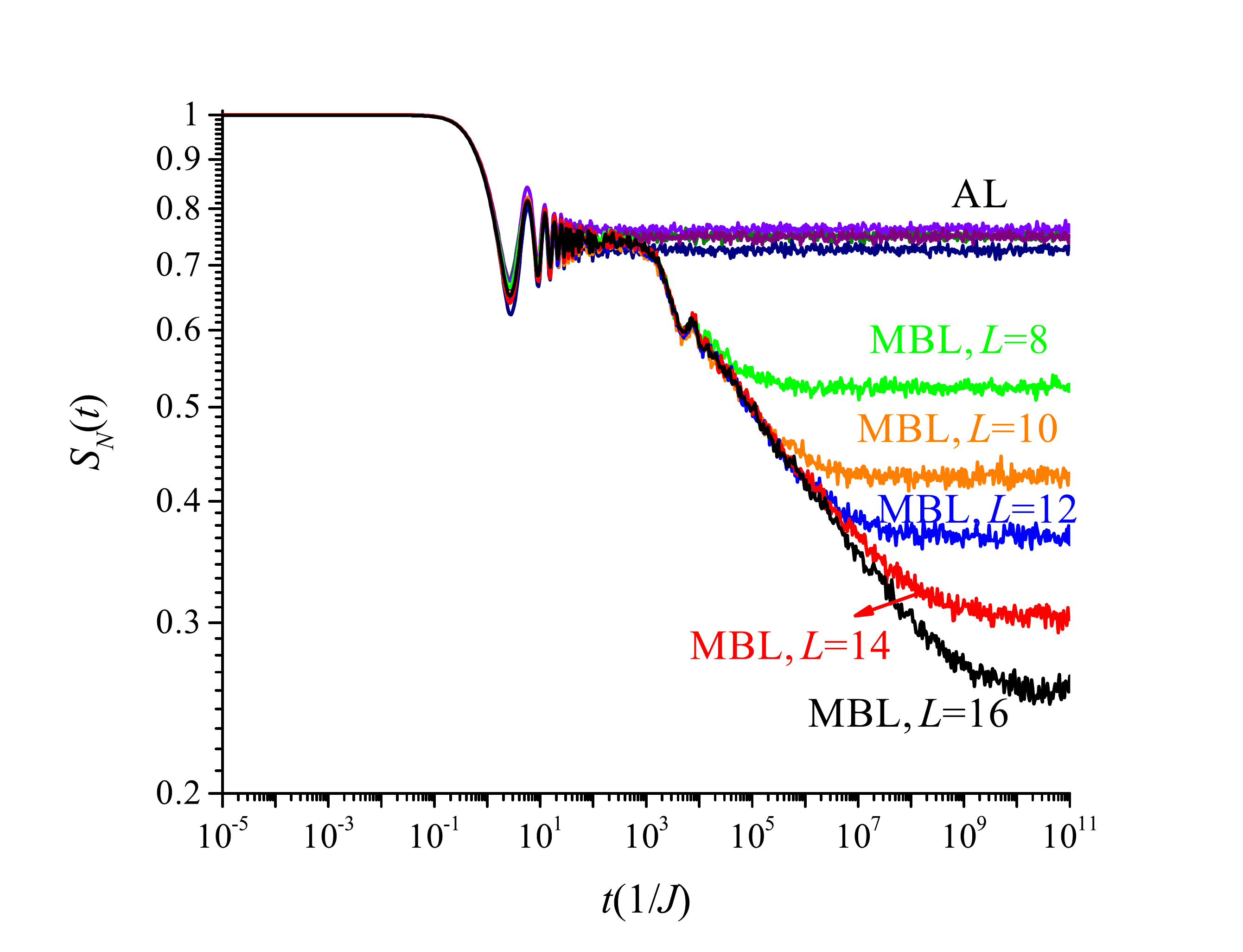}\caption{Logarithmic negativity s a function of time for disorder strength
$h=5$ and interaction strength $\Delta=0J (10^{-3}J)$ for AL (MBL)
phases. The results for AL phases with different spin chain length
$L$ are almost overlapping each other, and the results for different
MBL phases are indicated in the figure.}
\end{figure}

\end{document}